\documentclass[twocolumn,10pt]{article}
\usepackage{graphicx}
\usepackage[dvips]{color}
\input{epsf.tex}
\newcommand{\bibi}{\bibitem}                                                  
\newcommand{\etal}{\it {et al.}}                                             
\newcommand{\prl}{Phys. Rev. Lett.}                                       
                                           
\newcommand{\prb}{Phys. Rev. B}

\newcommand{\half}{\frac {1}{2}}                                             
                                             
\newcommand{\beq}{\begin{equation}}                                           
\newcommand{\eeq}{\end{equation}\noindent}                                  
\newcommand{\beqr}{\begin{eqnarray}}                                          
\newcommand{\eeqr}{\end{eqnarray}\noindent}                                   
\newcommand{\vQ}{\bf Q}                                                      
\newcommand{\vr}{{\bf r}}                                                     
\newcommand{\vk}{{\bf k}}

\newcommand{\cd}{c^{\dag}} 
\newcommand{\bd}{b^{\dag}} 
\newcommand{\hd}{h^{\dag}}

\newcommand{\noin}{\noindent}                                                 
\begin{document} 
\onecolumn
\title{Metallic conduction and superconductivity in the pseudogap phase}
\author{Sanjoy K. Sarker \\                                                   
Department of Physics and Astronomy \\                                        
The University of Alabama, Tuscaloosa, AL 35487 \\}                           
\maketitle    
\begin{abstract}
We analyze the $t-J$ model on a square lattice using bosonic spinons and fermionic holons. Spinons are paired into singlets, which condense below a temperature $T^*$. Metallic conduction and $d$-wave superconductivity result from separate, sublattice-preserving, holon hopping processes which originate from coupling with the condensate. Holons form a charge Fermi liquid, becoming incoherent (confined) above $T^*$. In the superconductor holons hop as pairs, reducing kinetic energy. The two-sublattice property is the glue that connects the three phases; its effect can be seen in various correlation functions. The theory can account for many features of the cuprate superconductors, including the origin of two-dimensional metallicity. 

  
\end{abstract}                                                                
\vspace{0.5 in}                                                              


\twocolumn

The normal state of a high-temperature superconductor is two-dimensional, and is not a conventional Fermi liquid. Moreover, a highly unusual \lq pseudogap' phase appears below a temperature $T^*$, superconductivity occuring at $T_c < T^*$ \cite{tim,lee1}. Anderson \cite{and} argued that the system is a Mott insulator (or, equivalently a quantum antiferromagnet), doped with holes; and proposed a metallic state in which spins are paired into singlets, or \lq valence bonds', and electrons split into separate charge and spin excitations, termed holons and spinons. The relevant microscopic model is the $t$-$J$ model on a square lattice, where $J$ is the antiferromagnetic (AF) interaction between neighboring spins, and $t$ describes nearest-neighbor (electron) hopping, such that no site is doubly occupied. The undoped insulator is in a mixed phase of a two-sublattice antiferromagnet, and a valence bond state. The physics is reasonably well described by a mean-field theory \cite{aro,sar1} in which the valence-bond state appears as a condensate of paired spinons - bosonic spin 1/2 neutral particles. It exhibits a two-sublattice character of its own (see below) --- singlet bonds connect spinons residing on opposite sublattices \cite{KRS}.

Much work has been done to extend the theory to the doped region \cite{bas,kot,lee1} with the assumptions that (i) moving holes rapidly destroy long-range AF order and (ii) the singlet condensate survives up to $T^*(x)$, which decreases with increasing hole concentration $x$. It is understood that there is no true long-range order associated with the singlets, $T^*$ is a crossover scale below which uniform magnetic susceptibility $\chi$ is strongly suppressed. The condensate is expected to account for the pseudogap, and also, it is hoped, that the singlets would acquire charge to form superconductig pairs. The physics of no double occupancy is accounted for by representing the electron as a composite object: $\cd _{i\sigma} = \bd _{i\sigma}h_i,$ where $\bd _{i\sigma}$ creates a spinon of spin $\sigma$ at lattice site $i$, and $h_i$ destroys a spinless holon, subject to the constraint: $h^{\dag} _ih_i + \sum _{\sigma}\bd _{i\sigma}b_{i\sigma} =  1$.  The Hamiltonian is given by 
\beq H = - t\sum _{ij,\sigma} \cd _{i\sigma}c_{j\sigma} ~-~ 
 2J \sum _{ij}{\rm A}^{\dag}_{ij}{\rm A}_{ij},  \eeq
where ${\rm A}_{ij} = \half \lbrack b_{i\uparrow}b_{j\downarrow} - 
b_{i\downarrow}b_{j\uparrow}\rbrack$ destroys a singlet. In both terms $i$ and $j$ are nearest neighbors (nn), and thus belong to opposite sublattices. The Hamiltonian has a local gauge symmetry since it preserves the number of holons plus spinons at each site. Usually, constraints are treated on the average, and the MF analysis is extended to obtain a generic metallic state, which is characterized by two competing order parameters, corresponding to {\em separately} broken gauge symmetry \cite{eli}. The first, $A_{ij} = <{\rm A}_{ij}>$ (where $<..>$ stands for averaging), characterizes the singlet condensate. The second, $D_{ij} = <\hd_jh_i>$, describes coherent hopping of holons on to nearest-neighbor (nn) sites, which does not preserve the two-sublattice property. The associated energy scale above which holons are incoherent is distinct from $T^*(x)$, with observable consequences. Many such MF states have been tried, some not consistent with the Mott state, but none has been found to be fully satisfactory.

Here we derive a different effective Hamiltonian by requiring that the Mott phase be described correctly. We assume that the singlet condensate evolves into the doped region with the same symmetry and statistics (bosonic spinons and fermionic holons \cite{aro,sar1,sar6}). Our analysis is based on two observations. (i) Even at the MF level, the \lq generic' metallic state (the spiral state \cite{sar2} in our case) is unstable \cite{sch,hu,sar4} towards insulating domain walls, or phase separation, in the entire parameter range of interest ($t/J \sim 3-4$, $x < 0.25-0.3)$. This absence of coherent inter-sublattice hopping has also been seen from single-hole calculations \cite{kane}. (ii) Hence, we set $D_{ij} = 0$, for nn (i,j). However, as we will see, for small $x$ the singlet condensate gives rise to two higher order sublattice preserving hopping processes, with symmetries determined by the symmetry of the the singlet phase. One leads to metallic conduction without additional symmetry breaking, by allowing holons to hop within a sublattice, and the other, discussed earlier \cite{sar5}, allows a singlet to hop onto a pair of holons, leading to d-wave superconductivity.  We find that for small $x$ the  theory is consistent with many properties of cuprates.

We first examine the symmetry properties in the Mott phase, using the MF solution \cite{aro}. In addition to singlets, spinons also condense, leading to AF order. For the singlet order parameter the choice (or, its gauge-equivalent) 
\beq A_{ij} = Ae^{i\half\vQ.(\vr_i-\vr_j)},\eeq 
where $\vr_i$ is a position vector (for unit lattice spacing), and $\vQ = (\pi,\pi)$, yields the correct state \cite{aro,sar1}. 
Eq. (2) leads to the spinon \lq\lq gap" function for the wave vector $\vk$, 
$\phi (\vk) = 4JA (\sin k_x + \sin k_y)$, which determines the properties of the system. Now, since $\phi (\vk) = \phi(\vQ - \vk)$, the singlet phase has a two-sublattice property of its own. Consider the pairing function $A_{ij} = \half <\lbrack b_{i\uparrow}b_{j\downarrow} - 
b_{i\downarrow}b_{j\uparrow}\rbrack>$ for any two sites $i$ and $j$. We find that $A_{ij}= - A_{ij}\cos(\vQ.(\vr _j - \vr _i))$. We will call this behavior {\em odd}, i.e., $A_{ij}$ is nonzero only if $i$ and $j$ are on {\em opposite} sublattices. Similarly, the spinon hopping function $B_{ij} = \half \sum _{\sigma} <\bd _{j\sigma}b_{i\sigma}>$ satisfies $B_{ij} = B_{ij} \cos (\vQ.(\vr _j -\vr _i))$. Hence, $B_{ij}$ is {\em even}, i.e., it is nonzero only if $i,j$ are on the {\em same} sublattice. These relations are intrinsic to
the singlet phase, and hold with or without long-range AF order, and at zero or finite $T$. They are gauge invariant, and have observable consequences. The spin-spin correlation function is given by ${\cal S}_{sp,ij} = <S^+_iS^-_j> =  - |A_{ij}|^2 + |B_{ij}|^2$, which, as expected, alternates in sign.   

For $x > 0$, we consider a uniformly charged state in the presence of a singlet condensate ($A \ne 0$), and use a renormalized perturbation expansion to study hopping up to one loop order, where the loop corresponds to the physical electron (i.e., the spinon-holon bubble). The structure of the theory is similar to that in ref. \cite{sar5}, except, now, $<\hd_jh_i> = 0$, for nn $i,j$. If $A = 0$ (e.g., above $T^*$), holons (and spinons) are confined or localized, as gauge symmetry is unbroken. The contributions to holon self energy are from confining (i.e, short-range) incoherent processes (similar to those found in single hole treatments \cite{kane}) in which holons propagate some distance, before returning. The singlet condensate breaks the gauge symmetry, giving rise to additional hopping terms, which can be derived by integrating out the spinons, and isolating the terms coupled to $A_{ij}$'s, and setting frequency $\omega = 0$\cite{sar5,sar3}. We assume that the short-range processes prevent AF order, and renormalize hopping amplitude to $t_{eff} < t$. Destruction of AF order leads to a nonzero spinon gap $\Delta _s$, so that $A_{ij}$'s are (exponentially) short-ranged. Then, for small $x$, we can derive a minimal holon Hamiltonian by retaining the short-range hopping terms, as follows. Other terms preserve the underlying symmetry, and thus, by continuity, will not change the physics qualitatively. 
 
When a hole hops from sublattice $a$ to $b$, it breaks a singlet and creates two spinons, costing an energy $\Omega \sim 2\Delta _s$. There are three ways to get rid of the energy. First, the hole hops back, which is confining. The other two lead to coherent motion, as shown below. 
\begin{figure}[htbp]
\centering
\includegraphics[height=2cm,width=8.0cm,angle=0]{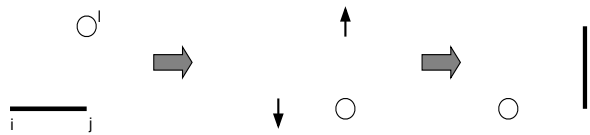}
\caption{One hole process. A hole hops from $l$ to $j$, breaking the singlet $(ij)$ denoted by the solid line. Here $i$ and $l$ are on sublattice $a$ and $j$ is on sublattice $b$. The hole then hops to $i$, and the singelet is reconstructed at $(jl)$}
\end{figure}   

\noindent A. {\em One-hole process --- metallic state}: The hole hops to another site on sublattice $a$, and the singlet is reconstructed on a different link [Fig. 1]. This yields an effective hopping term 
\beq -  \frac {2t_{eff}^2}{\Omega}\sum _{ijl} A^*_{jl}A_{ij}h^{\dag}_lh_i(1 - h^{\dag}_jh_j), \eeq
where $\hd$ creates a {\em renormalized} holon. We can replace the gauge invariant density $(1- h^{\dag}_jh_j)$ by its average value $1-x$. Using Eq. 2 for $A_{ij}$, we obtain the Hamiltonian
\beq H_{h} = \sum _{ij} t_{h,ij}h^{\dag}_jh_i, \eeq
which describes coherent holon propagation within the same sublattice, the sublattices are connected by a backflow of {\em singlets}. 
Here $t_{h,ij} = \mp t_h$ for $i,j$ next-nearest neighbors along $(1,\pm 1)$; $t_{h,ij} = - t_h/2$  for next-next-nearest neighbors, and $t_h = 4t_{eff}^2A^2(1-x)/\Omega$. Then the holon energy is  
\beq \epsilon_h(\vk) =  - 2t_h + 2t_h(\sin k_x + \sin k_y)^2. \eeq
The holon band (hence, metallic conduction) appears as soon as $A \ne 0$ without additional breaking of gauge symmetry. Eq.(3) is similar to hopping in short-range RVB models \cite{KRS}, provided one allows singlets to condense with appropriate symmetry. 

Let us consider correlation functions for {\em renormalized} particles, which clearly preserve the two-sublattice
property. Using $\epsilon_h(\vk) = \epsilon_h(\vQ - \vk)$, we find $D_{ij} =  <\hd_jh_i> = D_{ij} \cos (\vQ.(\vr _j -\vr _i))$, and thus is even. This leads to following results. (1) The magnetic correlation function has the same symmetry as in the Mott phase. (2) The electron hopping amplitude $P_{ij,\sigma} = <\cd _{i\sigma}c_{j\sigma}>= -B_{ij}D_{ij}$. Hence it is even, and decays exponentially, reflecting non-Fermi liquid behavior of the electron. Indeed, the corresponding electron Green's function (bubble) is incoherent (has no poles). Then the momentum distribution function satisfies: $n_c(\vk) = n_c(\vQ -\vk)$. (3) Let $\rho _i = \hd_ih_i - <\hd_ih_i>$ measure the excess hole density. Then the charge structure factor is given by: ${\cal S}_{ch,ij} = <\rho _i\rho _j> =  - |D_{ij}|^2,$ for $i \ne j$, and ${\cal S}_{ch,ii} = x(1 - x)$. Hence, it is even, and has the long-range oscillatory structure of a metal. In $\vk$ space we find: ${\cal S}_{ch}(\vk)  = {\cal S}_{ch}(\vQ - \vk)$. This is shown in Fig. 2. In contrast, ${\cal S}_{ch}(\vk)$ of an ordinary metal increases from zero at $\vk = 0$ and becomes a constant for $q > 2k_F$. In our case, an image of the behavior near $\vk = 0$ appears near $\vk = \vQ$.
\begin{figure}[htbp]
\centering
\includegraphics[height=6cm,width=6.0cm,angle=0]{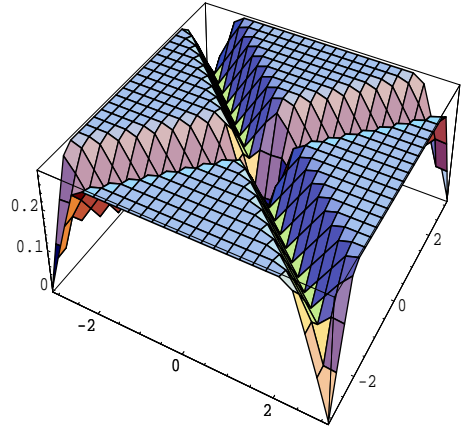}
\caption{Charge structure factor, calculated approximately. Note the symmetry between  $\vk = 0$ and $\vk = (\pi,\pi)$, which is distributed among the four corners of the Brilloin zone. The valley region is elevated relative to $\vk = 0$. In a normal metal, the structure near $(\pi,\pi)$ is absent.}
\end{figure}  
These properties are mostly hidden since the experiments probe the bare correlations which are dominated by incoherent processes that do not preserve the two-sublattice property. The best candidate is ${\cal S}_{ch}(\vk)$ since holon motion is coherent. The experimental ${\cal S}_{ch}(\vk)$ would no longer vanish at $\vQ$, but there will still be a dip. 

B. {\em Two-hole process -- superconductivity}: The system can also relax if a second hole hops from sublattice $b$ to $a$, and the singlet is reconstructed [Fig 3]. This yields 
a term $  -  t_s\sum _{ij;lm} A^*_{ml}A_{ij}\hd_i\hd_jh_lh_m, $  
where $t_s = 4t_{eff}^2/\Omega$, which describes hopping of a singlet, accompanied by the backflow of a  holon pair. This is the low $x$ form of the interaction derived earlier \cite{sar5}, but here the normal state is different. Using Eq.(2), we obtain,
\beq H_{h,int} = -t_0 \sum _{ij;lm}{\rm F}^{\dag}_{ij}{\rm F}_{ml}, \eeq 
where $t_0 = t_sA^2$, and ${\rm F}^{\dag}_{ij}= \hd_i\hd_j$ creates a holon pair on the link ${ij}$. The order of the indices is important, and follows from the symmetry (Eq. 2).
\begin{figure}[htbp]
\centering
\includegraphics[height=2cm,width=8.0cm,angle=0]{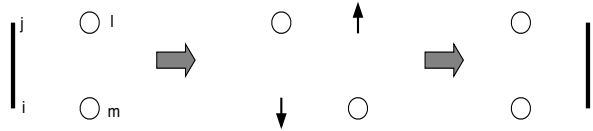}
\caption{Two hole process. A hole hops from $l$ to $j$, breaking the singlet $(ij)$, a second hole hops from $m$ to $i$, and the singlet is reconstructed at $(ml)$.}
\end{figure}  
Evidently kinetic energy is lowered if two holons hop as a pair. Pairs, being bosonic, will condense, leading to $F_{ij} = <{\rm F}_{ij}> \ne 0$. Let ${\cal C}_{ij} = 
<(c_{j\downarrow}c_{i\uparrow} - c_{j\uparrow}c_{i\downarrow})/2>$ denote the pairing order parameter for electrons. Then, ${\cal C}_{ij} = - A_{ij}F^*_{ij} \ne 0$, giving rise to superconductivity below $T_c \le T*$. 

Solution of the resulting mean-field problem depends on the symmetry of $F_{ij}$. Now, $F_{ij} = - F_{ji}$ (Fermi statistics). For a uniform system, $|F_{ij}| = F_0$, but the phases along $x$ and $y$ can be different. We can choose $F_{ij} = \pm iF_0$ along $\pm x$  and $F_{ij} = \pm i\alpha F_0$ along $\pm y$ direction, with $\alpha = e^{i\theta}$. Then, $\Delta _h(\vk) = 2t_0 F(\vk)$ is the holon gap function, where
\beq F(\vk) = 2F_0 (\sin k_x + \alpha \sin k_y). \eeq  
The choice of $\alpha = \pm 1$ leads to ${\cal C}_{x} = \pm {\cal C}_{y}$, corresponding to s-wave (d-wave) symmetry for the electron pair wave function. A numerical solution shows that $\alpha = -1$, (i.e., $d$-wave) yields the largest $F_0$, and hence the largest condensation energy. The origin of this result can be seen from the gap equation itself, which for real $\alpha$ and $T = 0$ is given by
\beq \frac{1}{t_0} = \frac{1}{N}\sum _k W(\vk) \frac{(\sin k_x + \alpha \sin k_y)^2}{E_{\vk}}
, \eeq
where $E_{\vk} = \lbrack (\epsilon _h(\vk) - \mu_h)^2 +  \Delta ^2_h(\vk)\rbrack ^{1/2}$ is the quasiholon energy, $\mu _h$ is the holon chemical potential and $W(\vk)$ is a sitably chosen cut-off function. The dominant contribution to the sum comes from the region where $|\epsilon _h(\vk) - \mu_h|$ is small, and the symmetry factor $|\sin k_x + \alpha \sin k_y|$ is large. As shown in Fig. 4, the holon Fermi surface is in the second and fourth quadrant, exactly where $\sin k_x + \alpha \sin k_y$ has maxima for $\alpha = - 1$ ($d$-wave) and vanishes for $\alpha = 1$ ($s$-wave)). Hence, $d$-wave always wins. Thus the symmetry is determined by the two-sublattice property of the normal and Mott phases!
\begin{figure}[htbp]
\centering
\includegraphics[height=6cm,width=6.0cm,angle=0]{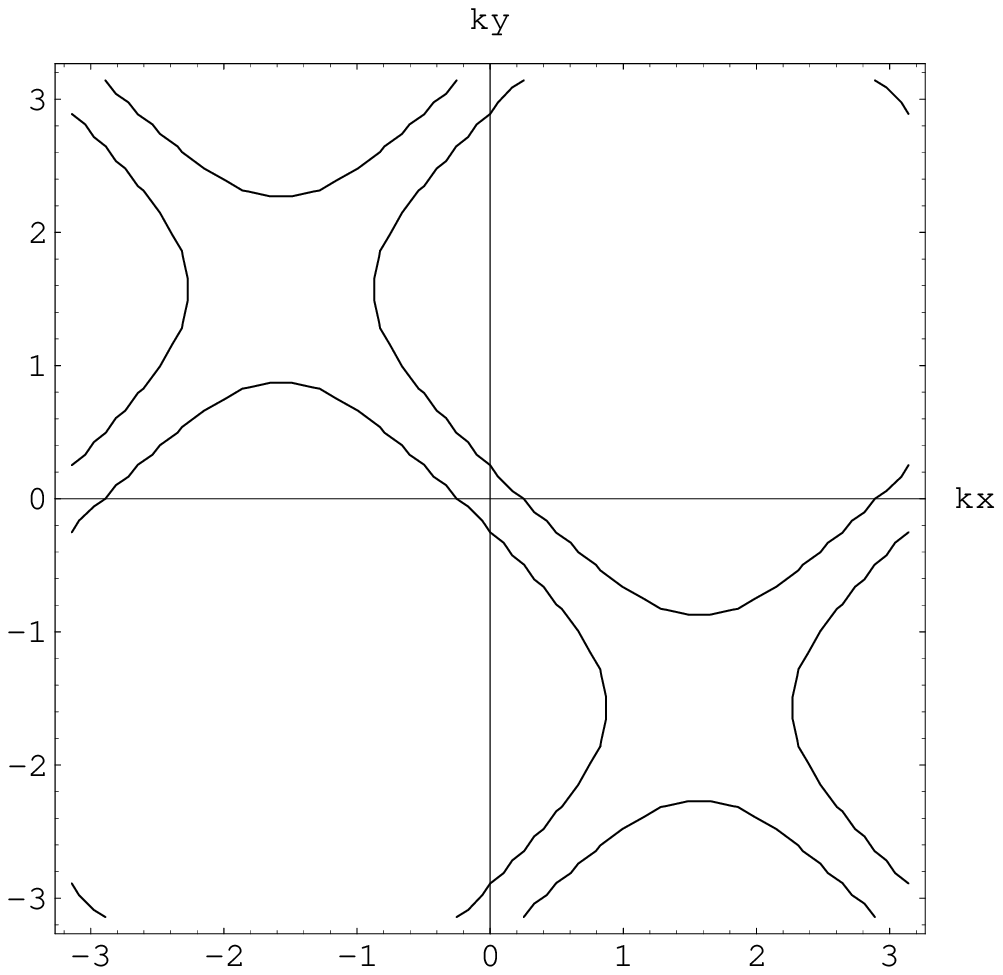}
\includegraphics[height=6cm,width=6.0cm,angle=0]{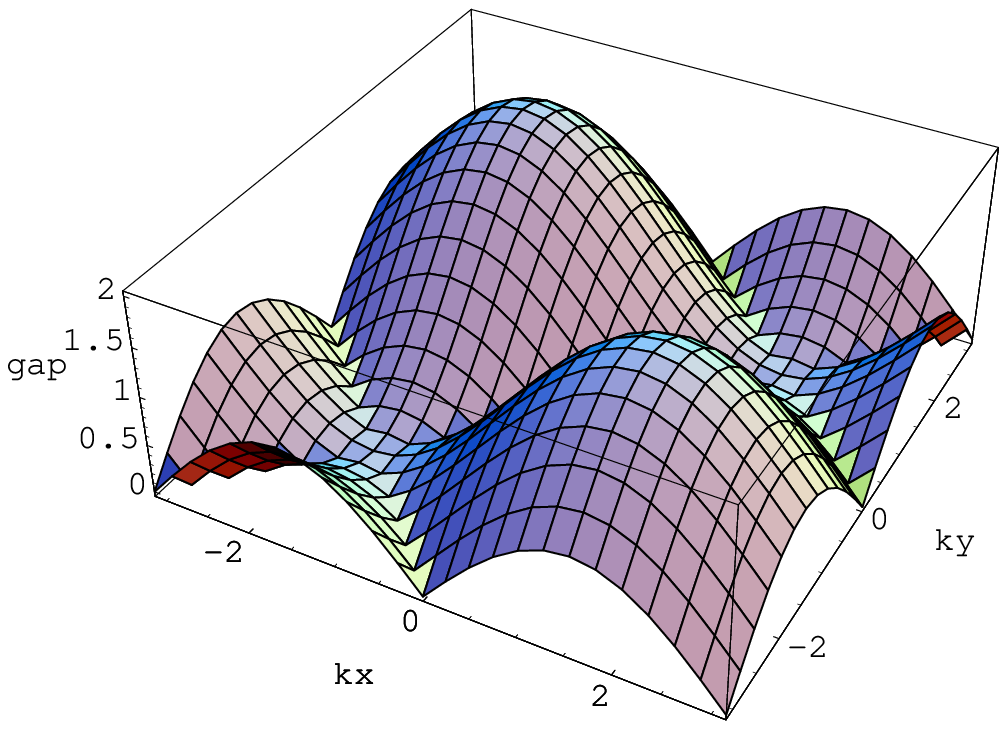}
\caption{Origin of $d$-wave symmetry. a) Holon Fermi surface. Holons live within the crossing strips centered at $\half(\pi,-\pi)$ and $\half(-\pi,\pi)$. b) The symmetry factor $|\sin k_x + \alpha \sin k_y|$. It has maxima on top of the FS for $\alpha = - 1$, satisfying the condition for maximum $F_0$, and hence maximum condensation energy. In contrast, the symmetry factor has minima in this region for  $\alpha = 1$ ($s$-wave). Under a gauge transformation both the Fermi surface and the symmetry factor move together to preserve these results}  
\end{figure}  
 
Since $F(\vk) = F(\vQ - \vk)$, superconducting state preserves the two-sublattice property. The holon pairing function for any two $i,j$  satisfies $F_{ij} = - F_{ij}\cos \vQ.(\vr_i - \vr _j)$. Hence the electron pairing function ${\cal C}_{ij}$ is nonzero only if $(i,j)$ are on opposite sublattices. The symmetries of $n_c(\vk)$ and spin-spin correlation function remain unchanged. The charge structure factor, however, picks up an additional contribution: ${\cal S}_{ch,ij} = |F_{ij}|^2 - |D_{ij}|^2$, and is no longer restricted to the same sublattice; but like the spin-spin correlation function, it oscillates in sign!

\noin C: {\em 2D vs 3D}: A key question in high-$T_c$ superconductivity is:  why is the normal state two dimensional, whereas superconductivity is not? This is easily resolved. Consider interlayer hopping of strength $t_z < t$. The corresponding exchange interaction $J_z = (t_z/t)^2J << J$. It follows that singlets form within the plane -- interalyer singlet excitations are gapped. Now, suppose a hole hops on to another layer (an electron hops back, by breaking a singlet). The effect is to create two unpaired spinons  -- one in each layer -- at a cost of $\Omega$. Now, there is no one-hole process for the system to relax, except by sending the hole back to the original plane. Hence the normal state is two dimensional. But superconductivity is three dimensional since it involves hopping of a singlet (or, equivalently, a pair of holons). This also explains the enhancement of $T_c$ due to interlayer hopping. 

\noin {\em Other Implications}:  For small $x$ (deep in the pseudogap regime), our simple theory has many implications for cuprates, some have been observed (as cited), and others can be taken as predictions. Here we list a few.  
Above $T^*$, the system is predicted to be in the {\em confined} phase: holons are localized due to gauge symmetry. There is no coherent charge carriers and, consequently, no Drude peak - only an incoherent background, with dc resistivity $\rho$ far exceeding the Mott limit \cite{tak}. Below $T^*$ holons become coherent, and form a \lq Fermi liquid' of concentration $x$, and effective mass determined by the band of width $W_h = 8t_h < W_0$, leading to a small plasma frequency. However, holon Fermi surface is not observable even in principle (lack of gauge invariance). For this system, (i) $\rho$ decreases rapidly below $T^*$ \cite{buch}, and becomes metallic (i.e, $< \rho _{Mott}$) at low $T$, with a residual impurity component and a $T^2$ term coming from fermion-fermion interaction. (ii) the optical conductivity $\sigma (\omega)$ has a Drude component, with an integrated area (spectral weight) $\propto x$ and a small plasma frequency \cite{orn}. It broadens by scattering as $T$ increases, and merges into the incoherent background above $T^*$ \cite{san,tak}. (iii) The Hall coefficient is positive and $\propto 1/x$, and independent of $T$ at low $T$ \cite{pad}. (iv) The holons contribute a $T$-linear term to heat capacity. (vi) The paramagnetic part of magnetic susceptibility $\chi$ is quenched below $T^*$, and vanishes as $T \rightarrow 0$ since spinons are gapped. Hence, total normal-state $\chi$ becomes more diamagnetic \cite{wang} at low $T$ due to the holon contribution. 
(vii) There are two \lq gaps' in the spin sector: a pseudogap associated with spinon pairing, which determines $T^*(x)$; and the actual spinon gap $\Delta _s$ which appears as soon as AF order is destroyed (for bosoinc spinons). From the MF theory we find,  $\Delta _s \sim c_s/\xi$, where $c_s$ is the spinon velocity, and $\xi$ is the AF correlation length. Therefore $\Delta _s$ (and the associated scale $T_s$) increases with $x$ with low doping. Then the physical electron {\lq bubble'} acquires a gap in the normal state. Whether $T_s$ can be identified with the Nernst scale remains to be seen. (viii) Superconductivity, however, appears via a weak-coupling induced pairing of holons -- therefore there are no {\lq preformed} Cooper pairs, and hence
we do not expect a large phase fluctuation region above $T_c$. (ix) Since both terms in the effective Hamiltonian arises from hopping, condensation energy ($\propto - t_s^2/t_h$), hence reduces kinetic energy as observed \cite{mole}. (ix) In the SC state, the Drude peak collapses to a delta function. 

For larger $x$, physical electrons will have to be taken into consideration \cite{sar5}. Our results do not change qualitatively when a small intra-sublattice hopping ($t^{\prime}$) term is included in the orginal model. A more elaborate method is required to describe gauge fluctuations and collective excitations, such as, nodal quasiparticles. However, the form of the effective Hamiltonian suggests the existence of a propagating spinon pair (singlet) excitation due to the backflow. This excitation carries neither spin nor charge, but is gapless, and hence would carry entropy and energy at low $T$. Useful discussions with N. Bonesteel, T. L. Ho, M. Ma, M. Randeria and C. Jayprakash is acknowledged. The author thanks the physics department of the Ohio State University, where part of the work has been done. The research at OSU was supported in part by a grant from NSF DMR-0426149 and a grant from DOE-Basic Energy Sciences, Division of Materials Sciences (DE-FG02-99ER45795).

\end{document}